\documentclass[twocolumn,runningheads]{svjour2}
\smartqed  % flush right qed marks, e.g. at end of proof
\usepackage{graphicx}
\journalname{Astrophysics and Space Science}
\begin{document}

\title{Spatial Orientation of Spin Vectors of Blue-shifted Galaxies}
%\subtitle{}

\titlerunning{}        % if too long for running head

\author{S. N. Yadav$^{\rm 1}$, \and B. Aryal$^{\rm 1}$ \and W. Saurer$^{\rm 2}$}

\institute    {First Author: S. N. Yadav \at
              $^{\rm 1}$Central Department of Physics, Tribhuvan University, Nepal\\
              \email{ysibnarayan@yahoo.com}           %  \\rew
\and          B. Aryal \at
              $^{\rm 1}$Central Department of Physics, Tribhuvan University, Nepal\\
              \email{binil.aryal@uibk.ac.at}           %  \\
\and          W. Saurer \at
              $^{\rm 2}$Institute of Astro- and Particle physics, Innsbruck
              University, Technikstrasse 25/8, A-6020 Innsbruck, Austria\\
              \email{water.saurer@uibk.ac.at}      }

\date{Received: 12 Dec 2015 / Accepted: .........}

\maketitle

\begin{abstract}
We present the analysis of the spin vector orientation of 5$\,$987
SDSS galaxies having negative redshift from $-$87.6 to $-$0.3
km$\,$s$^{-1}$. Two dimensional observed parameters are used to
compute three dimensional galaxy rotation axes by applying
`position angle--inclination' method. We aim to examine the
non-random effects in the spatial orientation of blue-shifted
galaxies. We generate 5$\times$10$^6$ virtual galaxies to find
expected isotropic distributions by performing numerical
simulations. We have written MATLAB program to facilitate the
simulation process and eliminate the manual errors in the process.
Chi-square, auto-correlation, and the Fourier tests are used to
examine non-random effects in the polar and azimuthal angle
distributions of the galaxy rotation axes. In general,
blue-shifted galaxies show no preferred alignments of galaxy
rotation axes. Our results support Hierarchy model, which suggests
a random orientation of angular momentum vectors of galaxies.
However, local effects are noted suggesting gravitational tidal
interaction between neighboring galaxies.


%\PACS{First \and Second \and More}
\end{abstract}

%________________________________________________________________

\section{Introduction}

It is commonly believed that the blue-shifted galaxies are
relatively nearby ones whose peculiar motion overcomes the Hubble
flow. All of the most distant galaxies (and indeed the
overwhelming majority of all galaxies) are red-shifted. According
to the conventional definition, the redshift of galaxies is the
sum of two terms: the isotropic cosmic expansion velocity and the
peculiar velocity owning to gravitational attraction by the
surrounding matter. In practice, determining the peculiar velocity
of a galaxy requires knowledge of both its observable radial
velocity relative to some reference system and the distance to the
galaxy determined independently of the radial velocity. According
to the linear theory of gravitational instability, the peculiar
velocities of galaxies are related to fluctuations in the mass
(Peebles 1980).

Burbidge \& Demoulin (1969) first observed IC 3258 with a
blueshift of $-$490 km$\,$s$^{-1}$. They give three possible
interpretations of their observations. First, IC 3258 is a member
of the Virgo cluster and has a very high velocity relative to the
average for the cluster. Second, IC 3258 is a field galaxy closer
to the Virgo cluster and its large velocity is just a random
motion. Third, IC 3258 has velocity because it has been ejected in
an outburst involving one of the radio galaxies in the Virgo
cluster. Several other blue-shifted galaxies appear in the
direction of the Virgo cluster.

By measuring the distance $d$ of a galaxy, one can obtain the
peculiar velocity of a galaxy $V_{pec} = V_{obs} - H_o\,d$, here
$H_o\,d$ is Hubble expansion velocity, $V_{obs}$ is observed
velocity of the galaxy. Since the Hubble expansion velocity is
small for nearby galaxies, the peculiar velocity could be
negative. Negative peculiar velocities are seen all over the
region around the Virgo cluster and this have long been seen as a
reflex of the pull of the cluster on us (Aaronson et al. 1982). We
live in the Local Supercluster, which is overdense part of the
Universe. So there is possibly the a local retardation of the
cosmic expansion or a net infall within this region. In another
example, an observer living on the outskirts of a large
concentration is also pulled towards the overdense part of the
clusters.

When the radiation propagates inside the collapsing body it is
blue-shifted. If this blueshift is greater than the  redshift
caused by the propagation of the radiation through expanding
universe, distant observer can detect the gravitational blueshift
from the collapsing object. Also the AGNs have blue-shifted
spectrum. Bian et al. (2005) studied the radial velocity
difference between the narrow emission-line components and of [O
III] $\lambda$ and H$\beta$ in a sample of 150 SDSS narrow-line
Seyfert$\,$1 galaxies. They found seven `blue outliers' with [O
III] blueshifted by more than 250 km$\,$s$^{-1}$. They interpreted
the blueshift as possible result of the outflowing gas from the
nucleus and the obscuration of the receding part of the flow by an
optically thick accretion disk and on the viewing angle.

Spatial orientation of angular momentum of blue-shifted galaxies
(SDSS) has not been studied, so we are interested to carry out the
spatial orientation of blue-shifted galaxies. An idea of the
origin of angular momentum of galaxies is very important to
understand the evolution of large scale structures of the
universe. This paper is organized as follows: in Sect. 2 we
describe the sample used and the method of data reduction. In
Sect. 3 we describe the methods, statistical tools and the
selection effects. Finally, a discussion of the statistical
results and the conclusions are presented in Sects. 4 and 5.

\section{Blue-shifted SDSS Galaxies}

We compiled a database of $5\,987$ blue-shifted galaxies from The
Sloan Digital Sky Survey seventh Data Release (SDSS DR7). All sky
distribution of blue-shifted galaxies is shown in Fig. 1a. The
inhomogeneous distribution of galaxies is because of the nature of
the survey. The distribution of blue-shifted galaxies is shown in
Fig. 1b. We found a linear relationship between the blue-shift and
logarithm of the number of galaxies ($-z$$\propto$log($N$)).
\begin{figure*}
\vspace{0cm} \centering
\includegraphics[height=5.0cm]{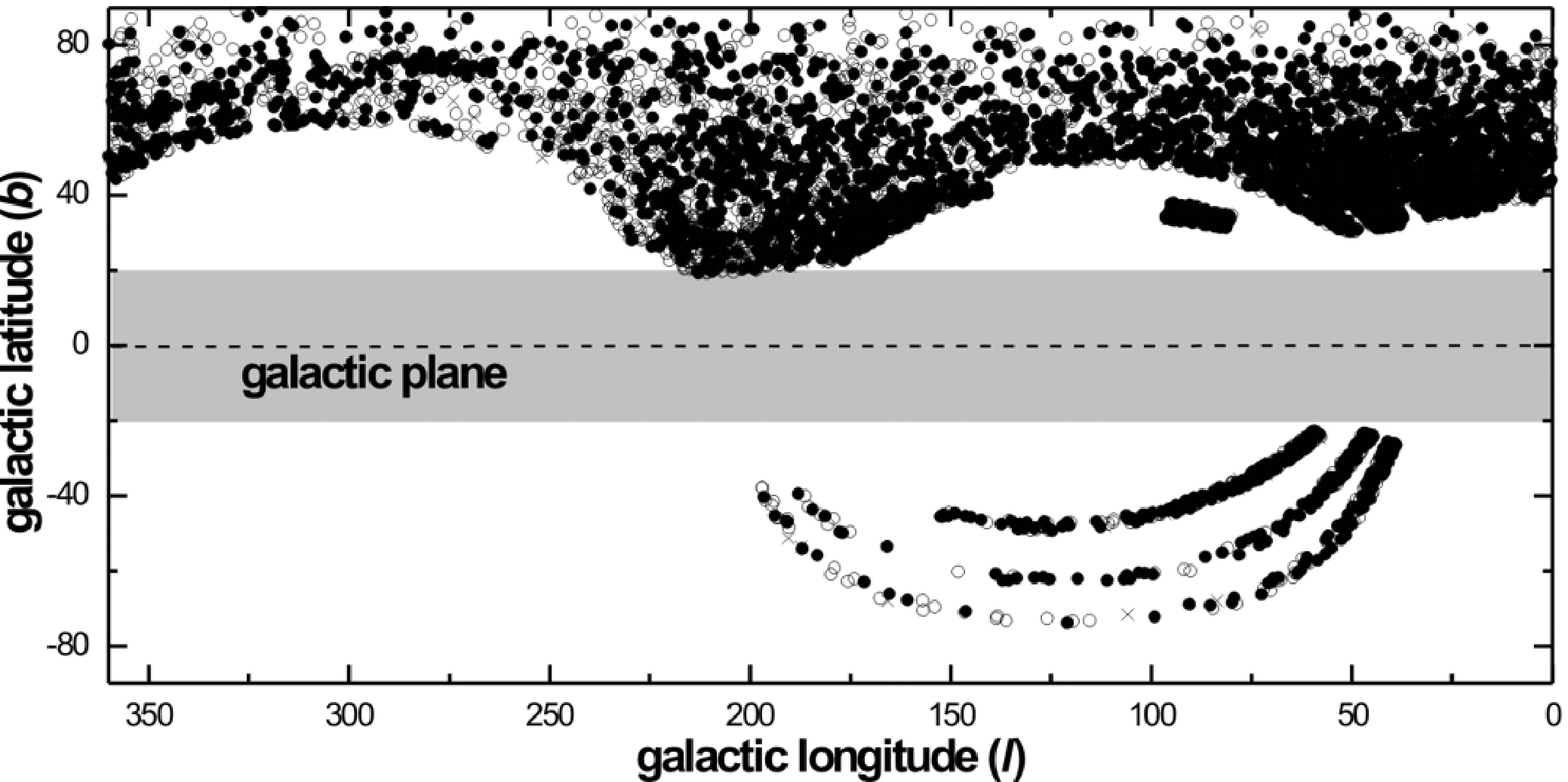}\\
\includegraphics[height=3.9cm]{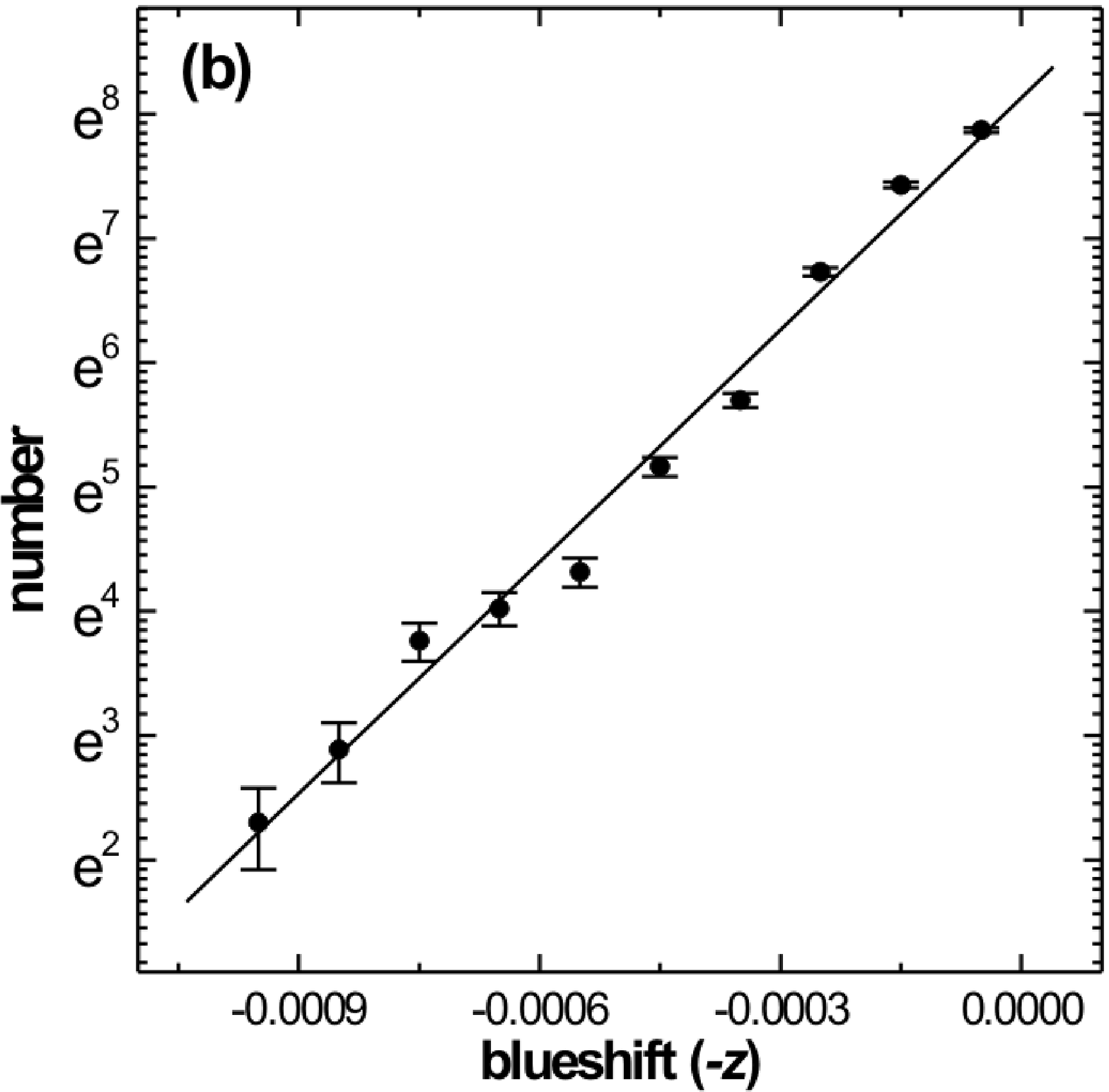}
\includegraphics[height=3.9cm]{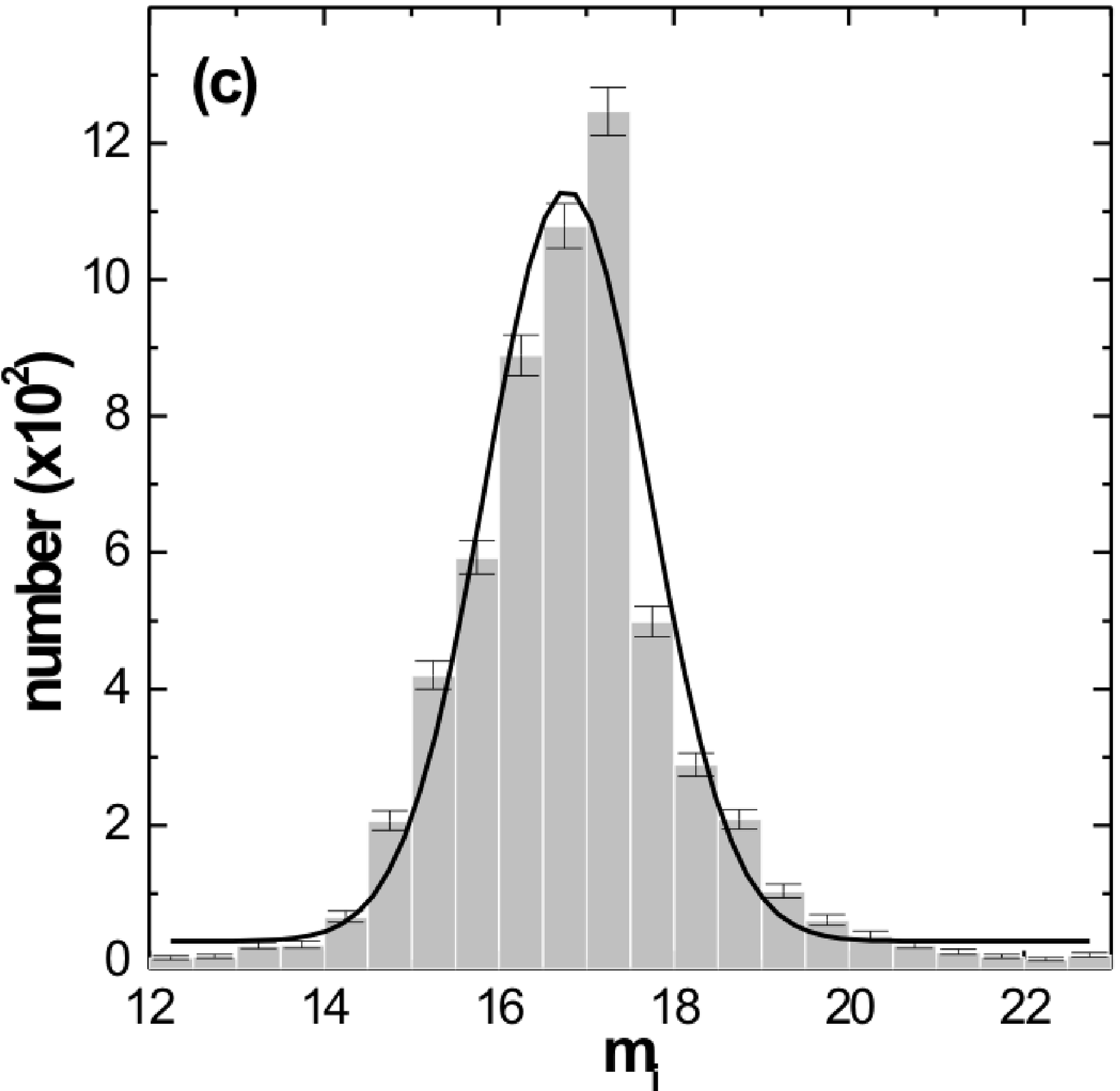}
\includegraphics[height=3.9cm]{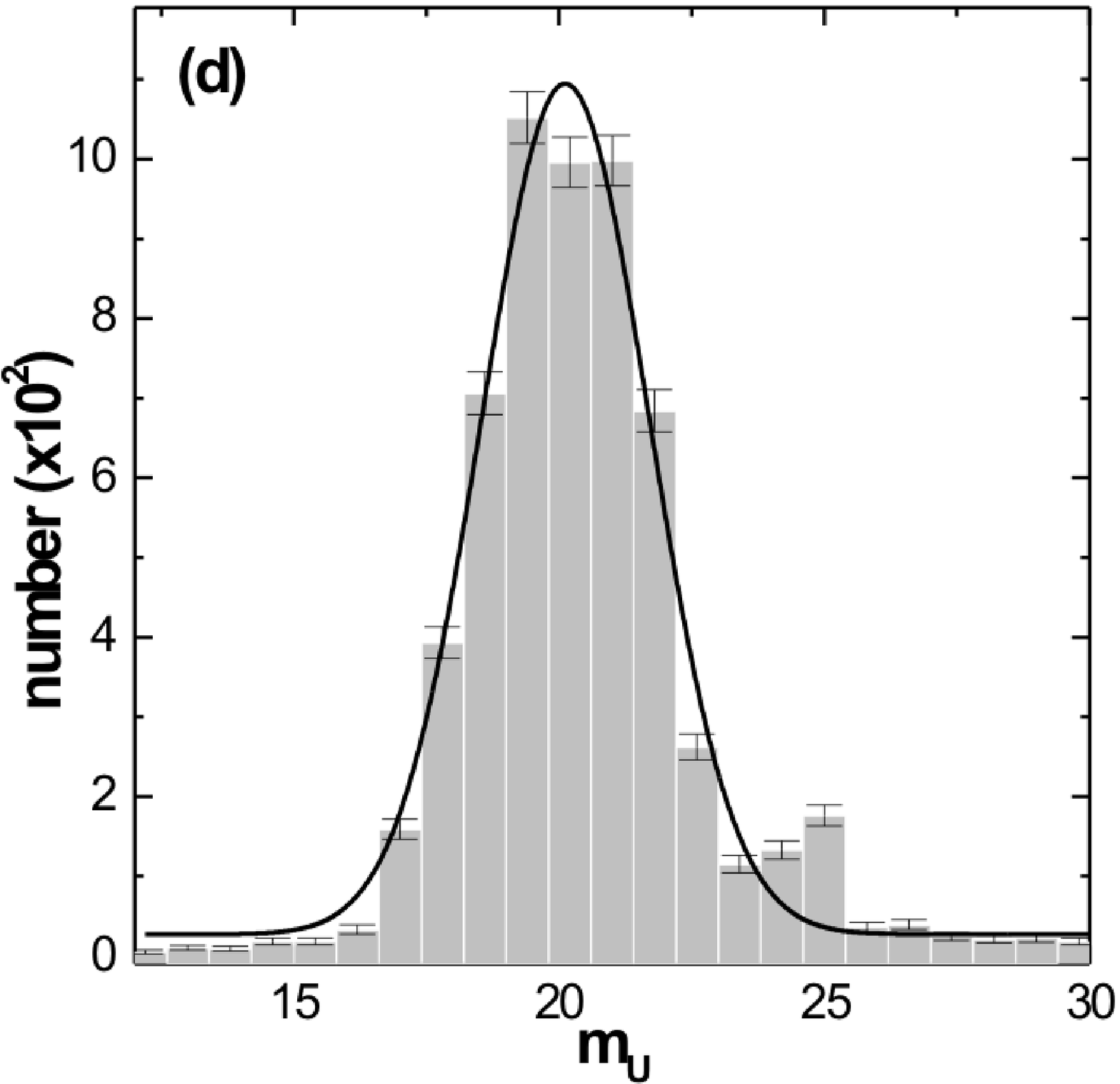}
\caption[]{ (a) All sky distribution of blue-shifted galaxies in
the galactic coordinate system. The grey-shaded region represents
plane of the Milky Way. The symbols solid circle, hollow circle
and cross represent small, moderate and strongly blue-shifted
galaxies (description in the text) (b) Blue-shift (c)
$i$-magnitude and (d) $u$-magnitude distribution of galaxies. The
solid line in (b) is the linear fit. The Gaussian fits are shown
by solid curves in (c) and (d). }
\end{figure*}
We have retained only those galaxies that have blue-shift ($-z$)
data at 95\% level of significance. This removed 569 galaxies form
the original data. Since blue-shift is found to be decreases
lineally with number (Fig. 1b), the remaining $4\,595$ galaxies
were classified into three bins by considering the bin size of
$1\times10^{-4}$. This resulted in three bins with number of
galaxies roughly in the ratio of $3:2:1$ in the largest, medium
and the smallest bins respectively. In the binning process,
galaxies that have very low and high blue-shift values were also
removed.

Since our galaxies are blue-shifted, their apparent magnitude
increases with time. In order to check the effect of blue-shift on
preferred alignments, we have chosen two extreme filters: infrared
($i$) and ultraviolet ($u$). The wavelengths of SDSS $i$ and $u$
filters are 7\,625 $\AA$ and 3\,543 $\AA$, respectively. The true
magnitudes of $i$ filter lies in the far-infrared and $u$ in the
visual bands. The study of far-infrared and optical activity in
the galaxy gives information regarding the early star formation
activity and the HII region, respectively. Fig. 1c,d shows the
magnitude distribution of near infrared and ultraviolet galaxies.
For both, Gaussian distribution fits well with the observed
distribution.

In order to find angular momentum vectors (or spin vectors, SV
hereafter), the diameters, position angles and positions of
galaxies should be known. We have compiled the database of
diameters and position angle of galaxies using SDSS survey.

\section{Method of analysis}

We follow the method suggested by Flin \& Godlowski (1986) to
convert two dimensional parameters (positions, diameters, position
angles of the DR7 SDSS blue shifted galaxies) into three
dimensional parameters (galaxy rotation axes in spherical polar
coordinates). The expected isotropic distribution for angular
momentum vectors or SVs of galaxies are determined by using the
method proposed by Aryal \& Saurer (2000). The observed and
expected distributions are compared with the help of three
statistical tests namely chi-square, auto-correlation and the
Fourier.

\subsection{Observed distribution: SVs of galaxies}
%___________________________________________________________________
\begin{figure}
\vspace{0.0cm} \centering
\includegraphics[height=6.5cm]{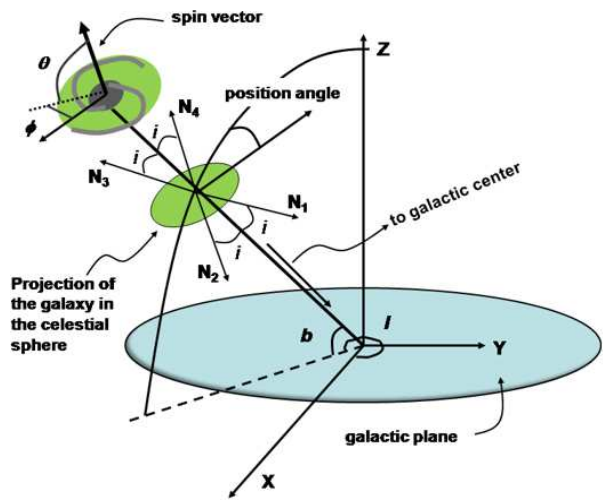}
\caption{Schematic illustration of $\theta$ (polar angle between
the SV of the galaxy and the reference plane) and $\phi$
(azimuthal angle between the projection of SV and the X-axis of
the reference plane). The galactic longitude ($l$) and latitude
($b$) are shown. For details: Flin \& Godlowski (1986) and Aryal
et al. (2008).}
\end{figure}
%___________________________________________________________________
The two-dimensional SDSS parameters (positions, position angles
and diameters) are converted into polar ($\theta$) and azimuthal
($\phi$) angles of galaxies using Flin \& Godlowski (1986). Our
blue-shifted galaxies have negative radial velocities probably due
to their larger peculiar velocity. These galaxies are mostly
nearby galaxies. Thus, it is convenient to use galactic coordinate
system as a physical reference plane. The formulae to obtain
$\theta$ and $\phi$ in are as follows (Flin \& Godlowski, 1986):
%____________________________________________________________
\begin{equation}
\begin{array}{l}
\sin \theta = - \cos i \sin b \pm \sin i \sin p \cos b\\
\end{array}
\end{equation}
%_____________________________________________________________
%_____________________________________________________________
\begin{equation}
\begin{array}{l}
\sin \phi = (\cos \theta)^{-1} [-\cos i\cos b \sin l\\
\,\,\,\,\,\,\,\,\,\,\,\,\,\,\,\,\,\,+ \sin i (\mp \sin p\sin b\sin l \mp \cos p\cos l)\\
\end{array}
\end{equation}
%_____________________________________________________________
where $l$, $b$ and $p$ are the galactic longitude, latitude and
position angle, respectively. The $i$ represents the inclination
angle, obtained using Holmberg's (1946) formula: cos$^2i$ =
[$(b/a)^2$--$q$$^2$]/(1--$q$$^2$) where $b/a$ is the measured
axial ratio and $q$ is the intrinsic flatness of disk galaxies.
The method of determination of intrinsic flatness of galaxies is
the same as in Aryal et al. (2013).

The above formulae show that there are two possible solutions for
a given galaxy. The normals ($N_{1}$, $N_{2}$, $N_{3}$, $N_{4}$)
shown in Fig. 2 can not be determined unambiguously, because we do
not know the side of the galaxy which is nearer/far to us, and the
direction of rotation. Thus, there are four solutions of the SV
orientation for a galaxy. We count all four possibilities
independently in our analysis.
%__________________________________________________________________
\subsection{Expected distribution: numerical simulation}

Aryal \& Saurer (2000) studied the effects of various types of
selections in the database and concluded that such selections may
cause severe changes in the shapes of the expected isotropic
distribution curves in the galaxy orientation study. Their method
has been applied by several authors in galaxy orientation studies
(Hu et al. 2006, Aryal et al. 2013 and the references therein).
Two kinds of selection effects are noticed in our database: (1)
inhomogeneous distribution of positions of galaxies, and (3) less
number of high inclination (edge-on) galaxies. These selection
effects are removed and the expected isotropic distribution curves
($\theta$ and $\phi$) are determined using the numerical
simulation method as proposed by Aryal \& Saurer (2000). For this,
a true spatial distribution of S of galaxies is assumed to be
isotropic. Then, due to the projection effects, $i$ can be
distributed $\propto$ $sin\ i$, latitude can be distributed
$\propto$ $cos\ b$, the variables longitude ($l$) and PA can be
distributed randomly, and formulae (1) and (2) can be used to
simulate (numerically) the corresponding distribution of $\theta$
and $\phi$. The isotropic distribution curves are based on
simulations including 10$^{7}$ virtual galaxies. The simulation
procedure is described in Aryal \& Saurer (2004). We perform
numerical simulation with respect to galactic coordinate system
systems. These expected isotropic distribution curves are compared
with the observed distribution.

\subsection{Statistical tests}

Our observed $\theta$ and $\phi$-distributions are compared with
expected isotropic distribution curves. For this comparison we
applied chi-square, autocorrelation and the Fourier (Godlowski
1993) tests. These tests are described in the appendix of Aryal et
al. (2007). These statistical tests are a proper method in our
case, because $\theta$ and $\phi$ are independent data. The
significance level is chosen to be 95\%, the null hypothesis is
established to be an equidistribution for the $\theta$ and $\phi$.

%________________________________________________________
\section{Results}

We have classified our database into six subsamples on the basis
of their redshift values and $u$$\,$\&$\,$$i$-magnitudes. Here we
discuss the distribution of the polar $(\theta)$ and azimuthal
$(\phi)$ angles of galaxy rotation axes in each subsamples. We
study the spatial orientation of SVs of galaxies with respect to
the galactic coordinate system. Any deviation from the expected
isotropic distribution will be tested using four statistical
parameters, namely the chi-square probability ($($P$>\chi^2)$),
auto-correlation coefficient ($($C/C$(\sigma))$), first order
Fourier coefficient ($\Delta_{11}\\/\sigma(\Delta_{11})$), and
first order Fourier probability ($($P$>\Delta_1)$). The conditions
for anisotropy are the following: P$(>\chi^2)$ $<$ 0.050,
C/C($\sigma$) and $\Delta_{11}$/$\sigma$($\Delta_{11}$) $>$ 1, and
P($>\Delta_1$) $<$ 0.150. These statistical limits were proposed
by Godlowski (1993, 1994). In the $\theta$-distribution, a
positive (negative) $\Delta_{11}$ suggests that the SVs of
galaxies tend to orient parallel (perpendicular) with respect to
plane of the Milky Way. In the $\phi$-distribution, a positive
(negative) $\Delta_{11}$ suggests that the SV projections of
galaxies tend to point radially (tangentially) with respect to
center of the Milky Way. Any `humps' or `dips' in the histogram
will be discussed. Here `hump' and `dip' are defined as having
more or less observed solutions than expected respectively.
%----------------------------------------------------------------
\begin{table*}
\centering \caption{Statistics of the polar $(\theta)$ and
azimuthal angle ($\phi$) distributions of galaxies in the six
subsamples. The first column lists the subsamples. The second and
third columns give the number of galaxies (N) in the subsample and
their blue-shift (-$z$, in km s$^{-1}$). The fourth and fifth
columns list the values of chi-square probability (P$(>\chi^2)$)
and auto-correlation coefficient (C/C$(\sigma)$). The last three
columns gives the first order Fourier coefficient
($\Delta_{11}/\sigma(\Delta_{11})$), the first order Fourier
probability (P$(>\Delta_1)$), and standard deviation of Fourier
amplitude ($\sigma(\Delta_1)$) respectively. }
\begin{tabular}{@{}cccccccc@{}}
subsample & N & $-z$ &P$(>\chi^2)$ & C/C$(\sigma)$ &
$\Delta_{11}/\sigma(\Delta_{11})$ & P$(>\Delta_1)$ &
$\sigma(\Delta_1)$ \\
\hline
polar angle\\
\hline
$i01$    &  1813  &   $<$28.5       & 0.781     & $-$0.1        & $-$0.4    & 0.939    & 0.023         \\
$i02$    &  1144  &   28.5-58.5     & 0.025     & $+$1.9        & $-$0.1    & 0.849    & 0.030         \\
$i03$    &  569   &   $\geq$58.5    & 0.023     & $-$1.7        & $-$0.7    & 0.773    & 0.042         \\
$u01$    &  1297  &   $<$28.5       & 0.644     & $-$1.0        & $+$1.2    & 0.404    & 0.028         \\
$u02$    &  839   &   28.5-58.5     & 0.896     & $-$0.5        & $-$0.2    & 0.931    & 0.035         \\
$u03$    &  438   &   $\geq$58.5    & 0.146     & $+$1.8        & $-$0.4    & 0.879    & 0.048         \\
\hline
azimuthal angle\\
\hline
$i01$    & 1813  &   $<$28.5        & 0.657     & $+$0.1        & $-$1.1    & 0.432    & 0.023         \\
$i02$    & 1144  &   28.5-58.5      & 0.107     & $-$1.4        & $-$2.1    & 0.090    & 0.030         \\
$i03$    & 569   &   $\geq$58.5     & 0.449     & $+$1.5        & $-$2.0    & 0.093    & 0.042         \\
$u01$    & 1297  &   $<$28.5        & 0.039     & $+$0.2        & $-$2.2    & 0.098    & 0.028         \\
$u02$    & 839   &   28.5-58.5      & 0.433     & $+$0.5        & $-$1.7    & 0.116    & 0.035         \\
$u03$    & 438   &   $\geq$58.5     & 0.860     & $+$0.3        & $+$0.1    & 0.747    & 0.048         \\
\hline
\end{tabular}
\end{table*}

\subsection{Polar angle distribution}

All three statistical tests suggest isotropy in the $\theta$
distribution of subsample $i$01 (Table 1). The galaxies in this
subsample are nearby blue-shifted (RV $<$ $-$28.5 km s$^{-1}$)
having $i$-magnitude in the range 12.45 and 32.08. In the
histogram, no significant humps or dips can be seen (Fig. 3a).
Thus, a random orientation of angular momentum vectors of galaxies
is found, suggesting hierarchy model of the structure formation as
suggested by Peebles (1969).

\begin{figure}
\centering \vspace{0.0cm}
\includegraphics[height=4.3cm]{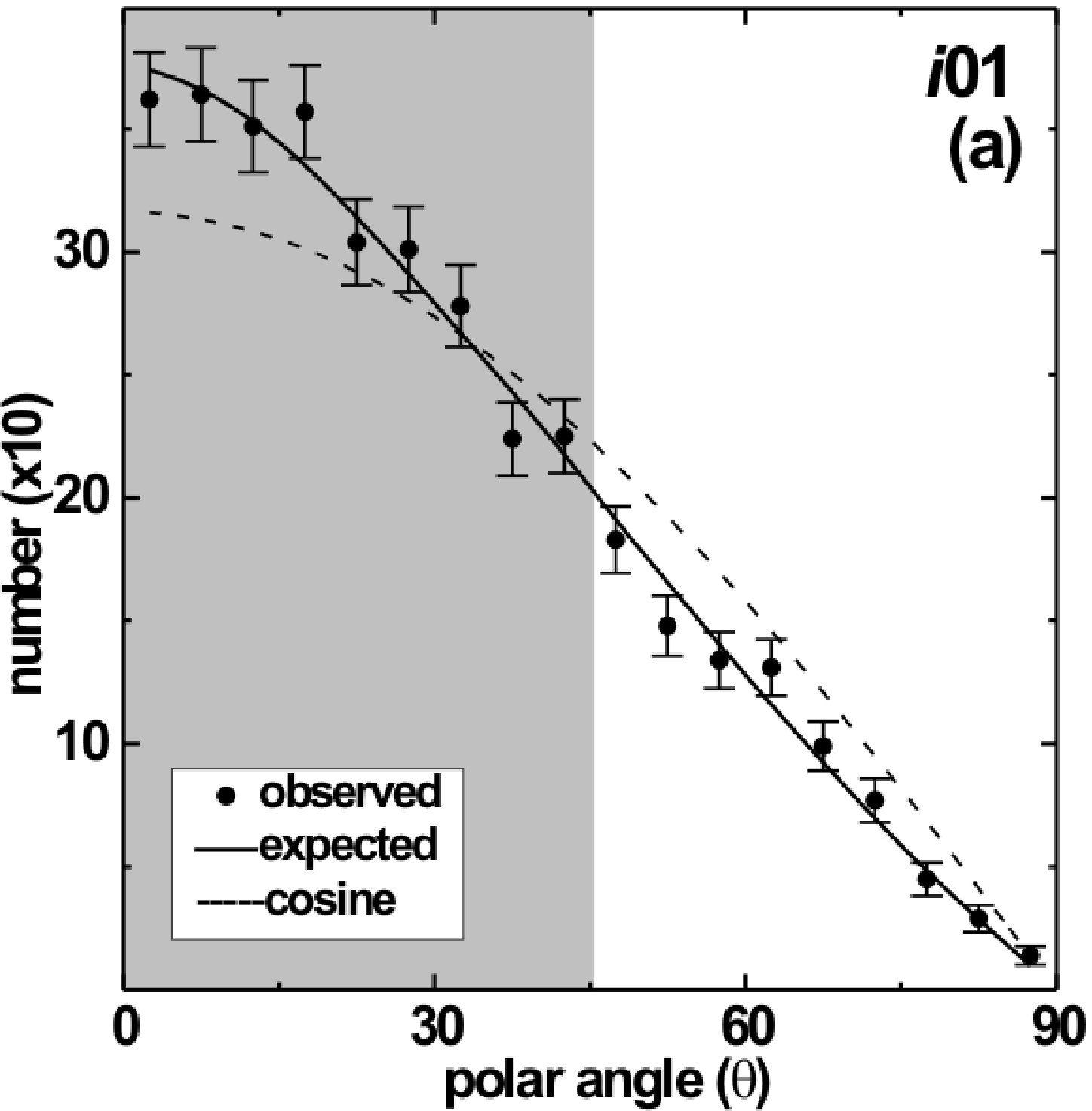}
\includegraphics[height=4.3cm]{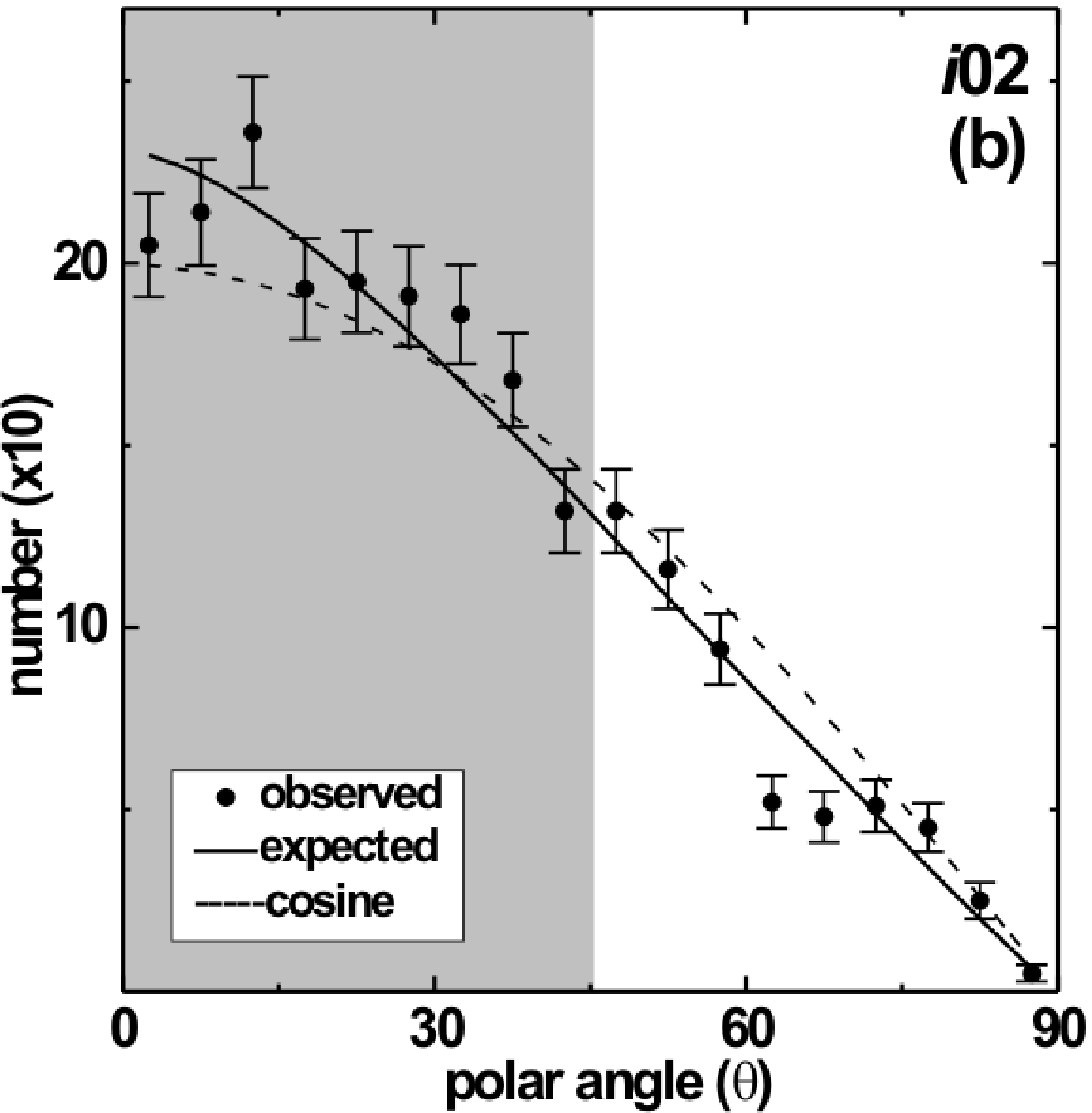}
\includegraphics[height=4.3cm]{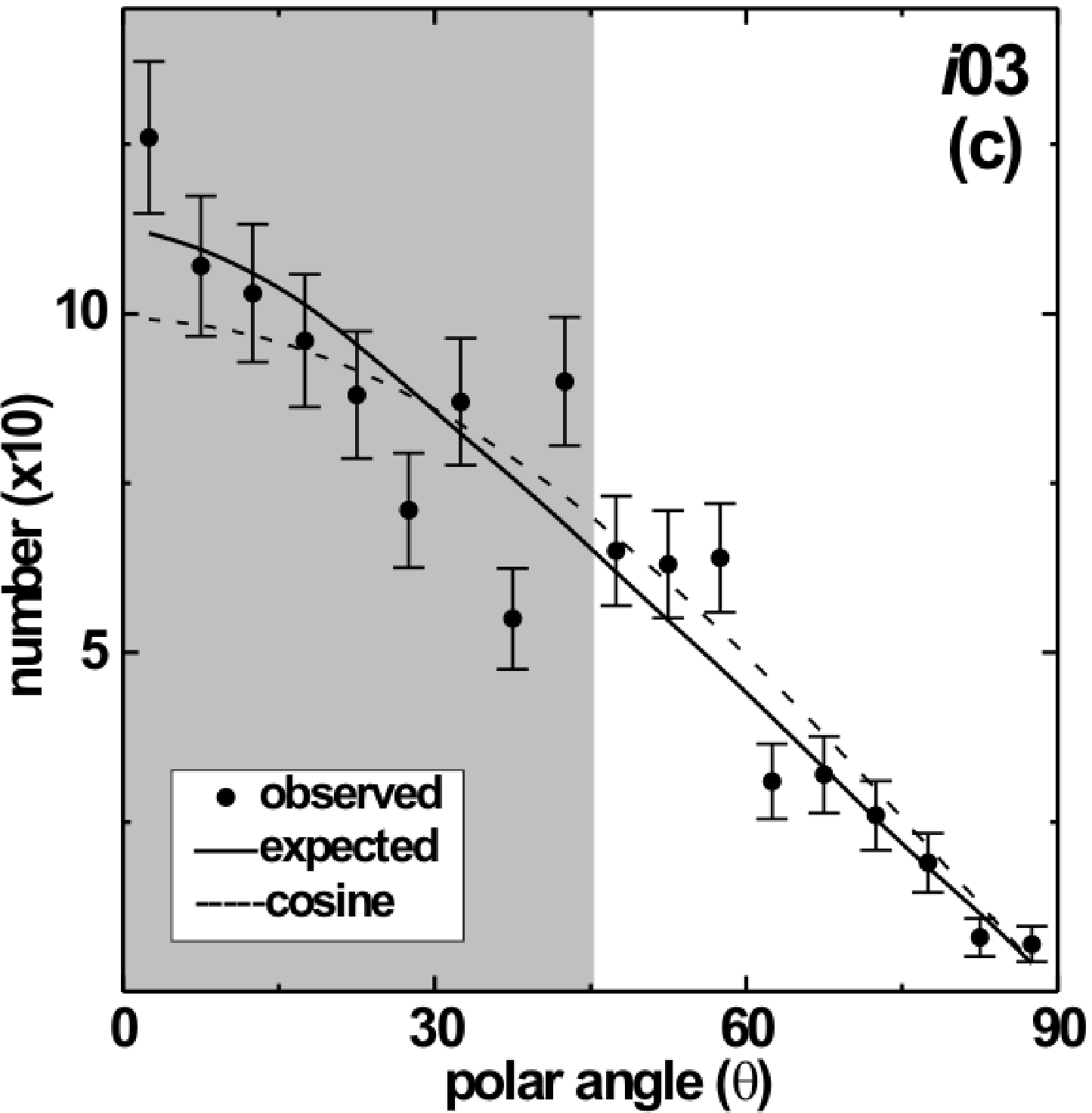}
\includegraphics[height=4.3cm]{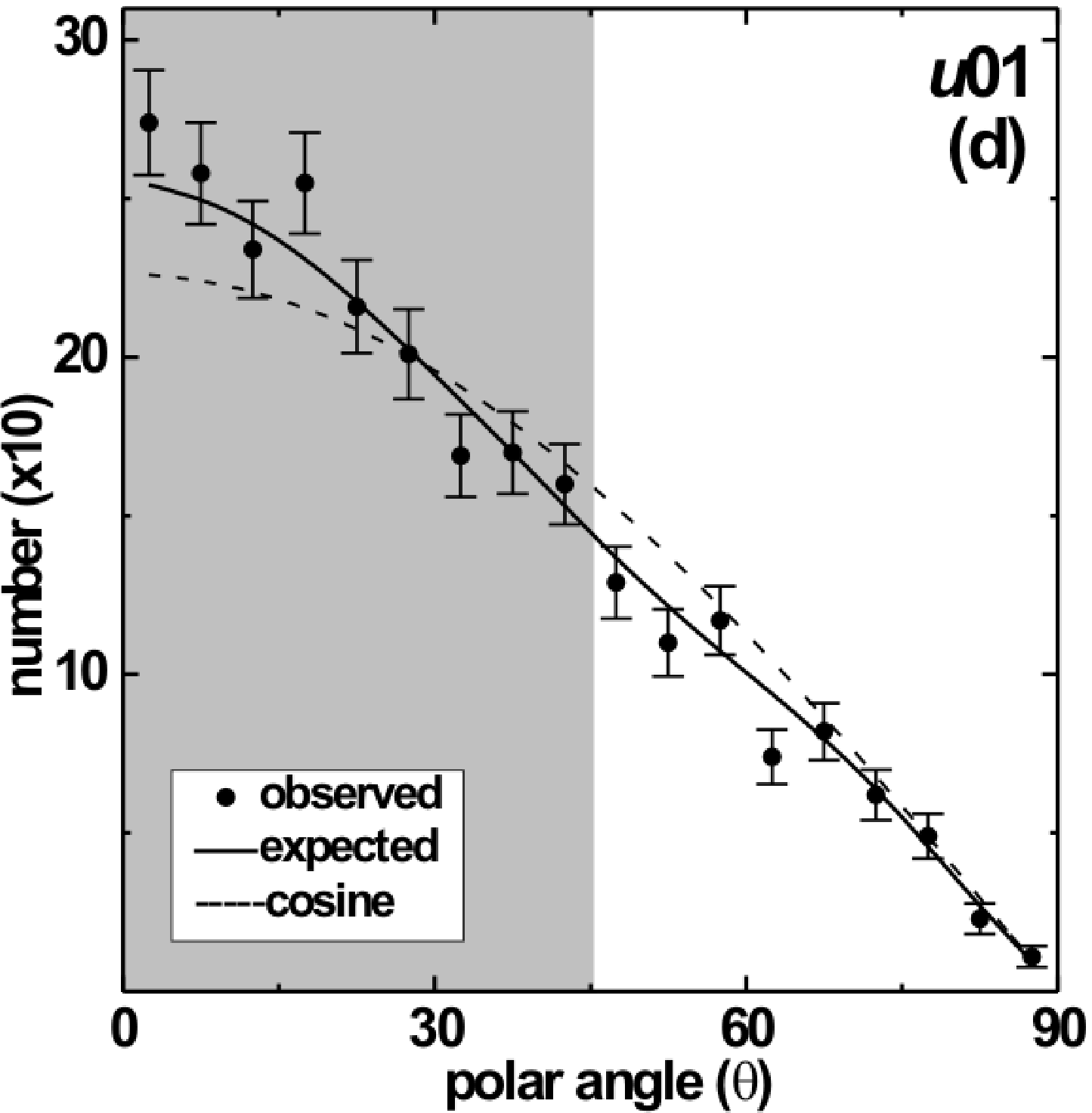}
\includegraphics[height=4.3cm]{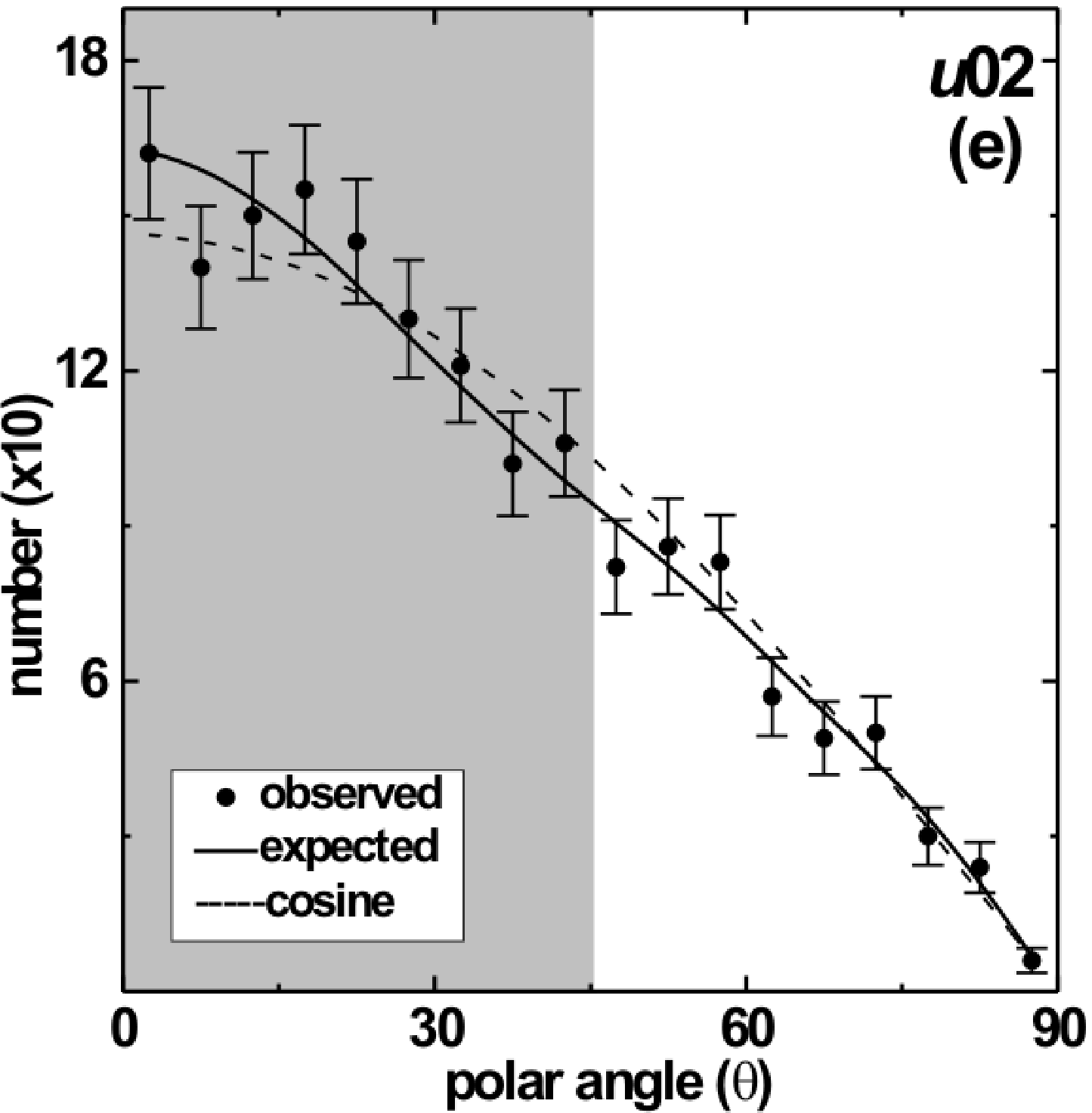}
\includegraphics[height=4.3cm]{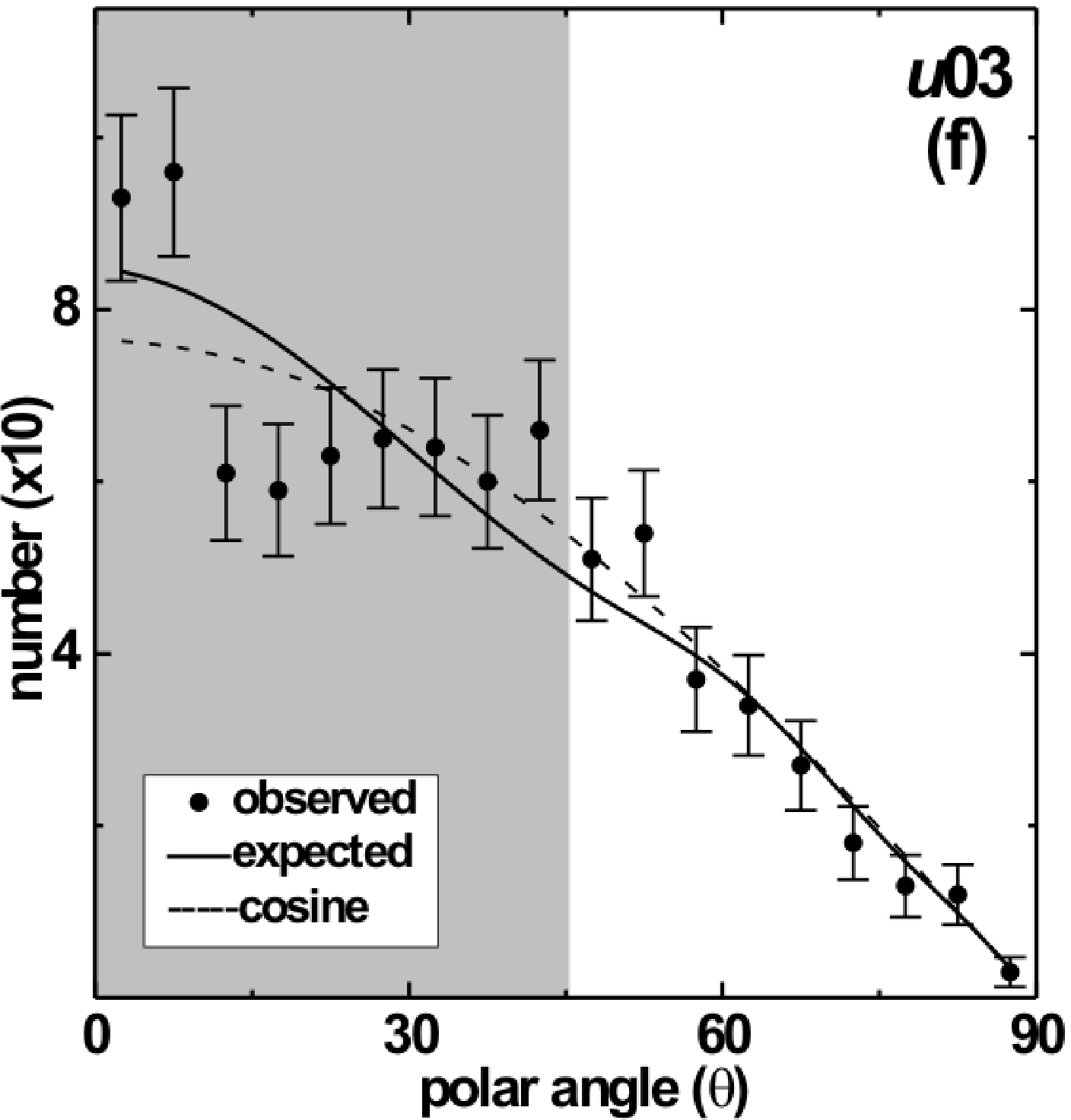}
\caption[]{The polar ($\theta$) angle distributions of blue
shifted SDSS galaxies in 6 subsamples. The solid circles with
$\pm$ 1$\sigma$ error bars represent the observed distribution.
The solid line represents the expected isotropic distributions.
The cosine distributions (dashed) are shown for the comparison.}
\end{figure}

The galaxies in the subsample $i$02 are moderately blue shifted
(RV: 28.5-58.5 km s$^{-1}$ and m${_i}$: 10.46 - 36.10). The
chi-square and the correlation tests show anisotropy (Table 1).
However, the first order Fourier probability and the first order
Fourier coefficient suggest isotropy. Since the observed
distribution follow the expected, we regard the Fourier test as
more reliable. There are small dips at 5$^{\circ}$, $\sim$
65$^{\circ}$, and a hump at 15$^{\circ}$ (Fig.3b). These are
possibly due to the binning effect.

Strong blue-shifted (RV: $\geq$58.5 km s$^{-1}$) SDSS galaxies
(m${_i}$: 10.82 - 42.23) are grouped in the subsample $i$03. In
the histogram, a very good agreement between the observed and
expected distributions can be seen. Similar to the subsample
$i$03, chi-square probability and correlation coefficient show
anisotropy, whereas first-order Fourier probability and the
first-order Fourier coefficient suggests isotropy (Table 1).
Several humps (0$^{\circ}$, 45$^{\circ}$, 60$^{\circ}$) and dips
(30$^{\circ}$, 35$^{\circ}$, 60$^{\circ}$) suggests the local
effect. Thus, the strong blue-shifted infrared galaxies are in the
process of grouping or subclustering because of the tidal or
gravitational shearing effect.

All three statistical tests support isotropy in $\theta$
distribution of $u$-magnitude low blue-shifted (RV $<$ 58.5 km
s$^{-1}$) galaxies. In Fig. 3d, no significant humps or dips are
found. A small hump at 20$^{\circ}$ and a dip at 60$^{\circ}$ are
because of the binning effects. These effects do not change the
statistics of the subsample $u$01. Hence, no preferred alignments
of SVs of galaxies in subsample $u$01 is noticed.

In subsample $u$02, all three statistics show isotropy in polar
angle ($\theta$) distribution. A small dip at 10$^{\circ}$ is
because of the binning effect (Fig. 3e). A random orientation of
SVs of galaxies is noticed in this subsample.

All statistics except correlation coefficient showed isotropy in
the polar angle distribution of high blue-shifted (RV $\geq$ 58.5
km s$^{-1}$) $u$-magnitude galaxies (subsample $u$03). The
correlation coefficient is found to be 1.8 ($>$1.5$\sigma$ level).
This is because of the humps at 10$^{\circ}$, 45$^{\circ}$,
55$^{\circ}$ and dips at 15$^{\circ}$, 20$^{\circ}$. These humps
and dips suggests the possibility of grouping or subclustering
because of the tidal or gravitational shearing effects between
comoving galaxies.

To sum up, the spatial orientation of blue-shifted galaxies is
found to be random, supporting hierarchy model (Peebles 1969) of
structure formation. It is found that the rapidly moving blue
shifted galaxies are in the process of large structure (galaxy
groups, subclusters, clusters, etc) formation. In addition,
magnitude is found to be independent of the preferred alignments.

\subsection{Azimuthal angle distribution}

All three statistics support isotropy in the azimuthal angle
distribution of low blue-shifted high magnitude infrared
(subsample $i$01) galaxies (Table 2). There is a hump at
$+$65$^{\circ}$ and dips at $-$5$^{\circ}$ and $-$55$^{\circ}$
(Fig.4a). These are due to the binning effect and hence do not
change the statistics. Thus, no preference is noticed for spin
vector projections of low blue-shifted infrared galaxies.

All statistics except $\chi^2$-test suggest anisotropy in the
$\phi$ distribution of the subsample $i$02. In the histograms for
the $\phi$-distribution of the subsample $i$02 (Fig.4b), humps at
$-$60$^{\circ}$, $+$30$^{\circ}$ and $+$90$^{\circ}$, and dips at
$-$50$^{\circ}$ to $+$20$^{\circ}$ can be seen. These humps and
dips lead this subsample to show anisotropy. The spin vector
projection of $i$02 galaxies is found to be directed tangentially
outwards with respect to the center of the Milky Way.
\begin{figure}
\centering \vspace{0.0cm}
\includegraphics[height=4.25cm]{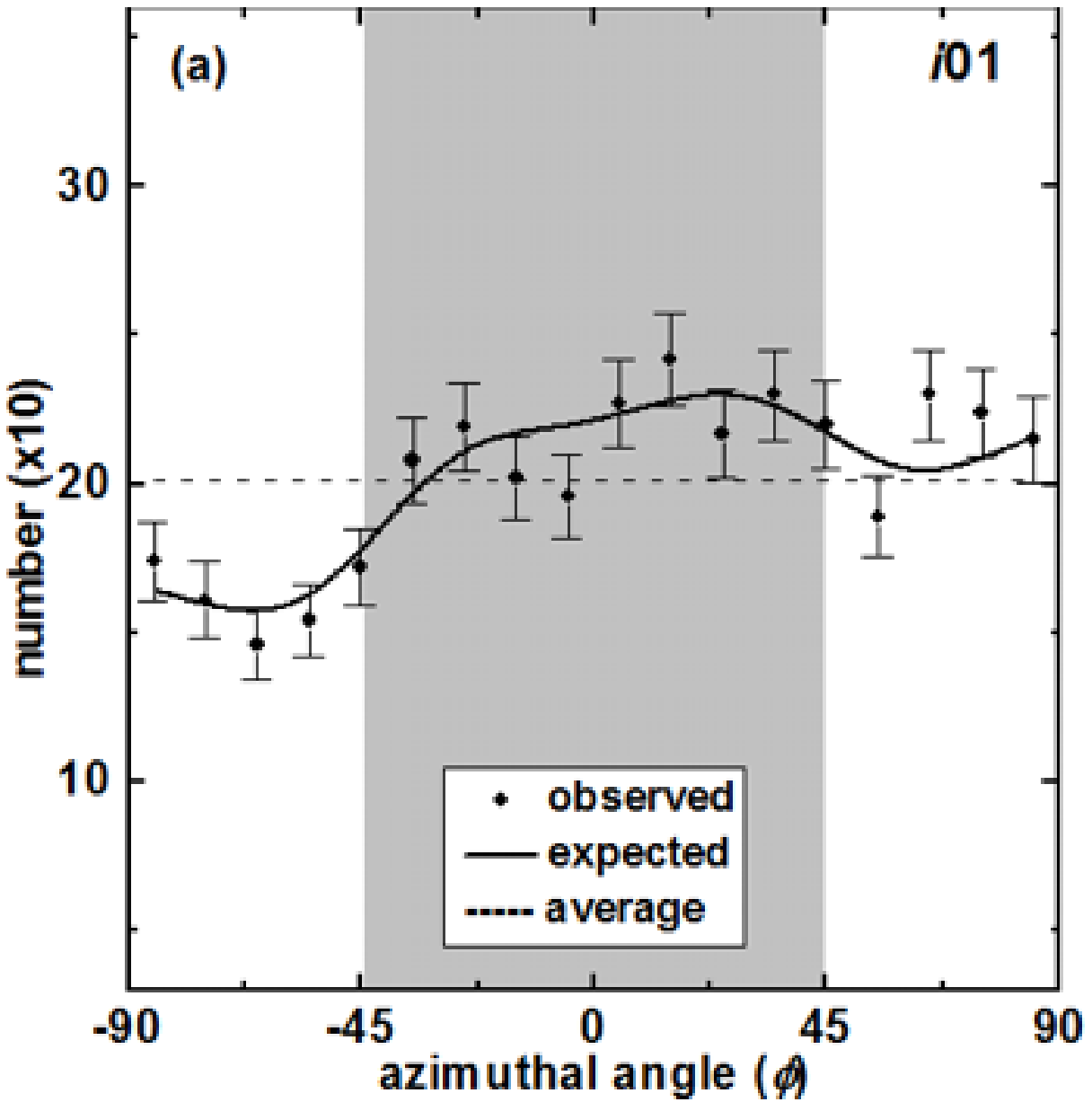}
\includegraphics[height=4.25cm]{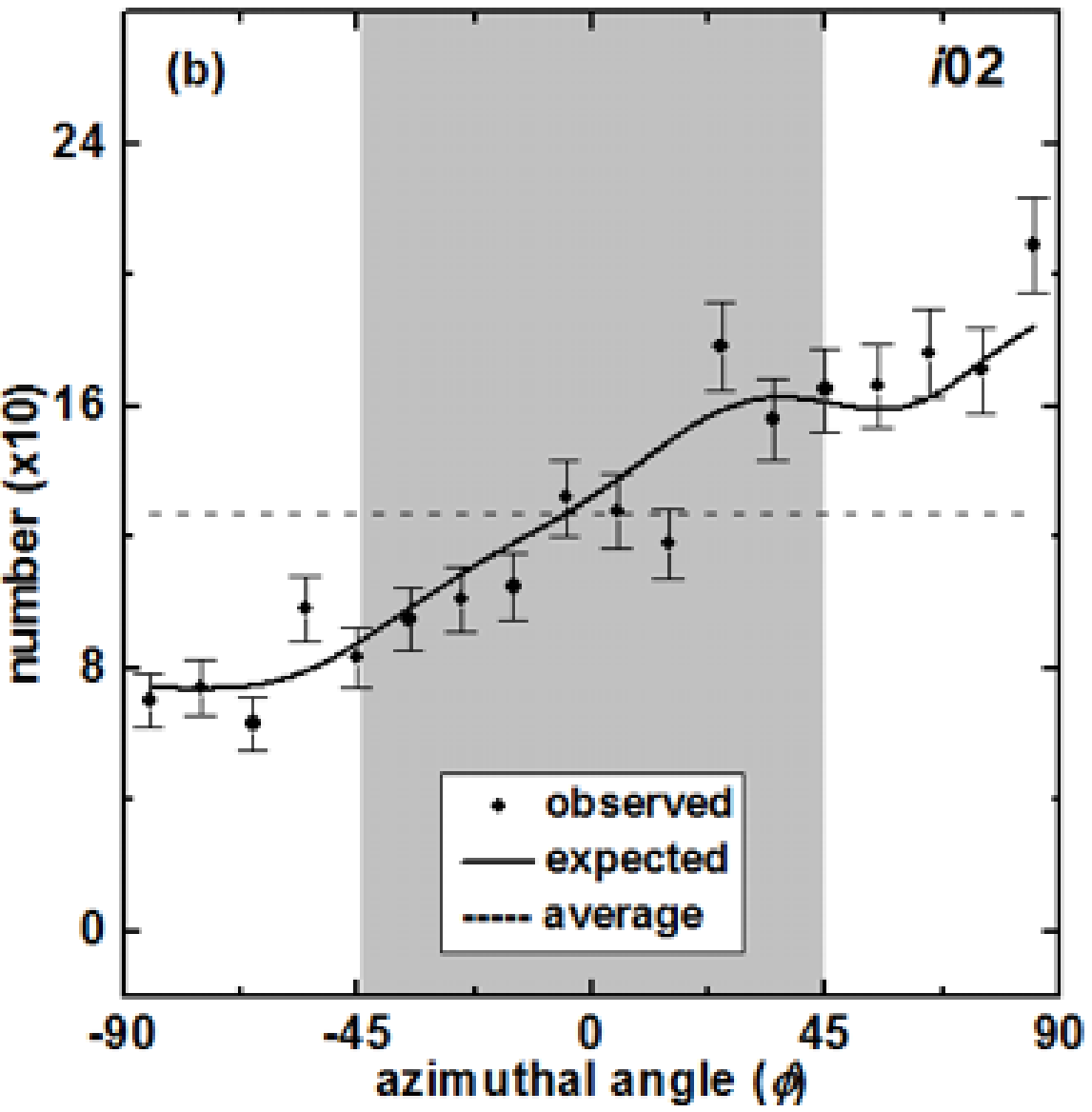}
\includegraphics[height=4.3cm]{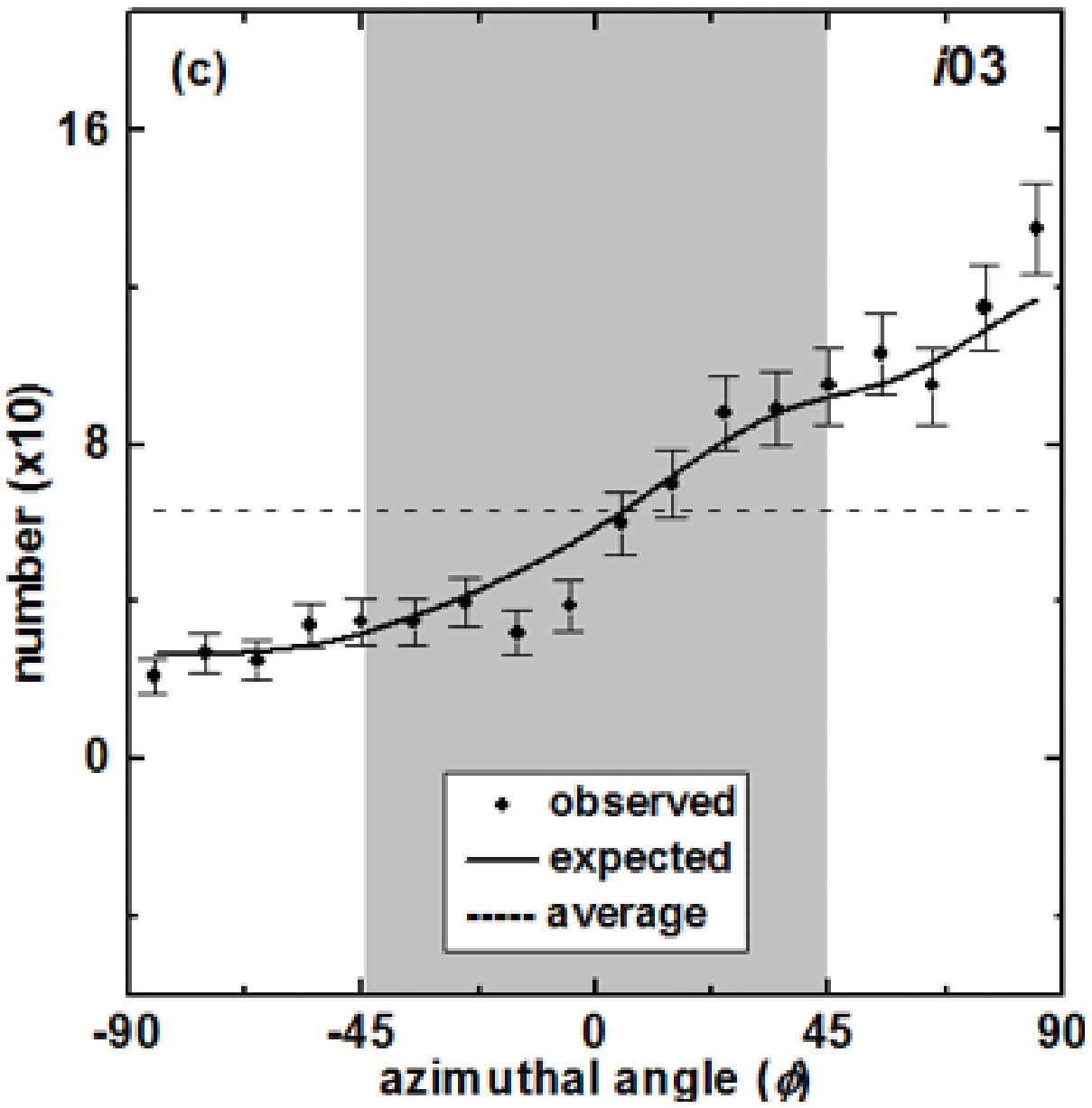}
\includegraphics[height=4.3cm]{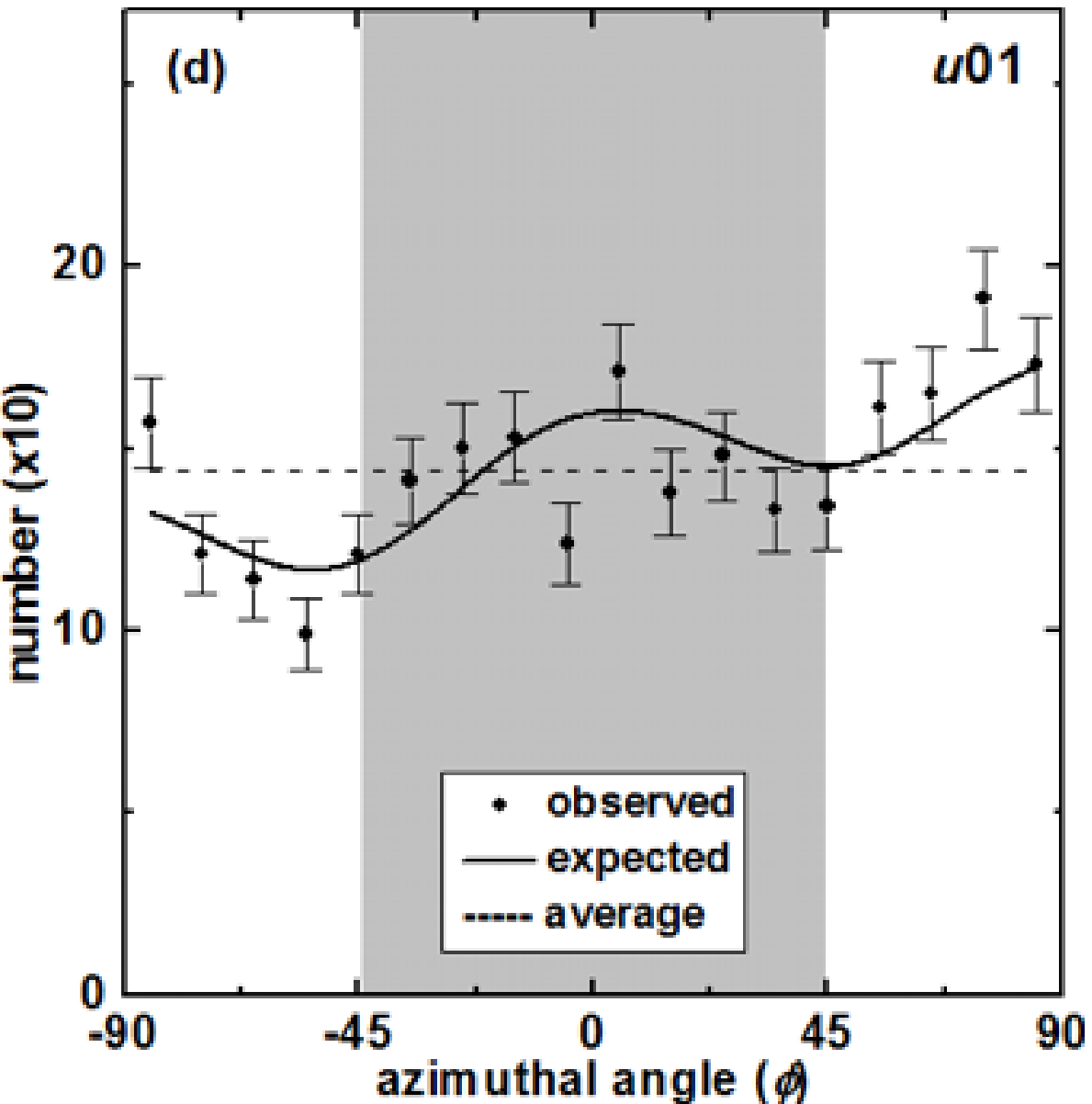}
\includegraphics[height=4.3cm]{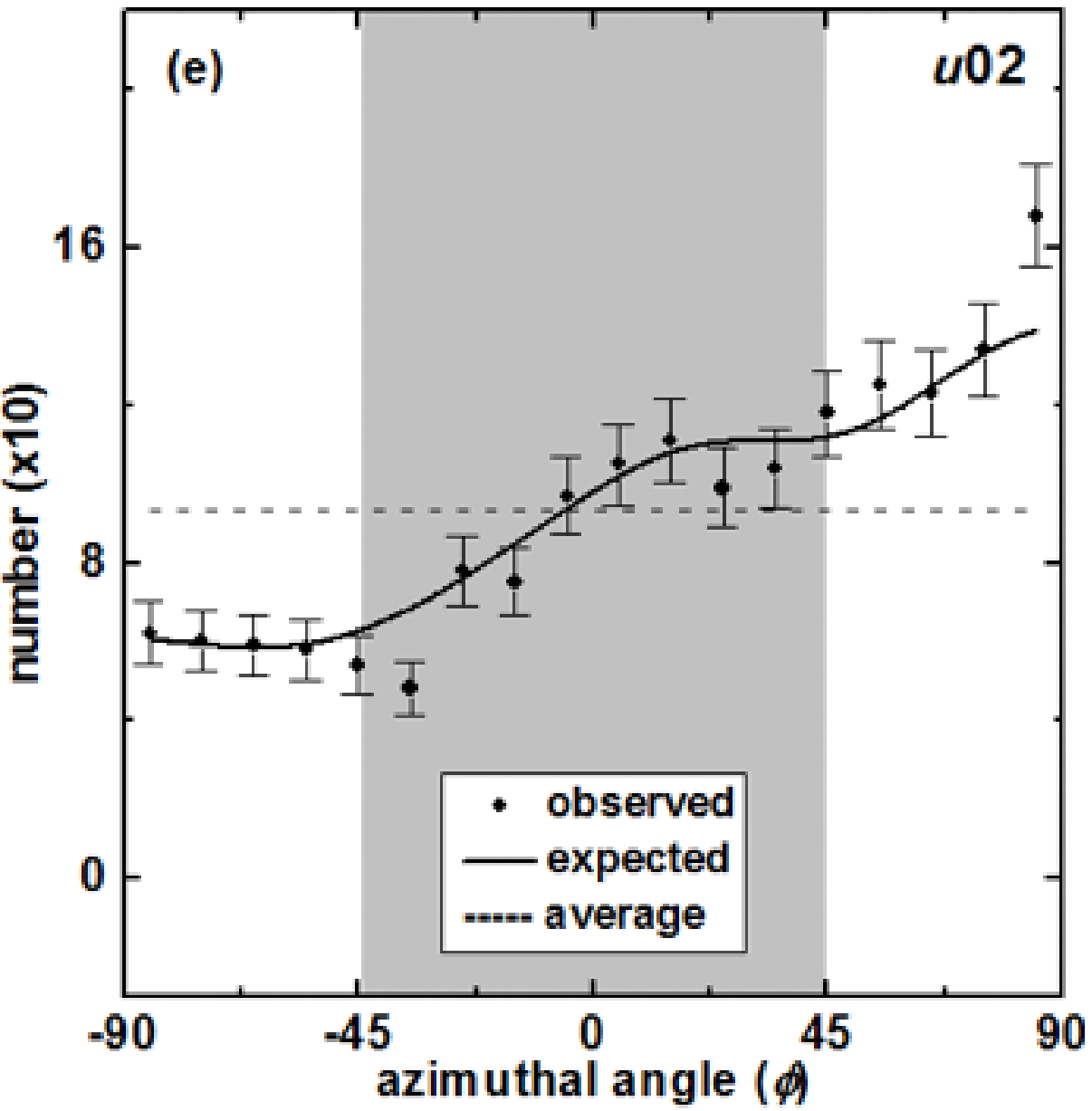}
\includegraphics[height=4.3cm]{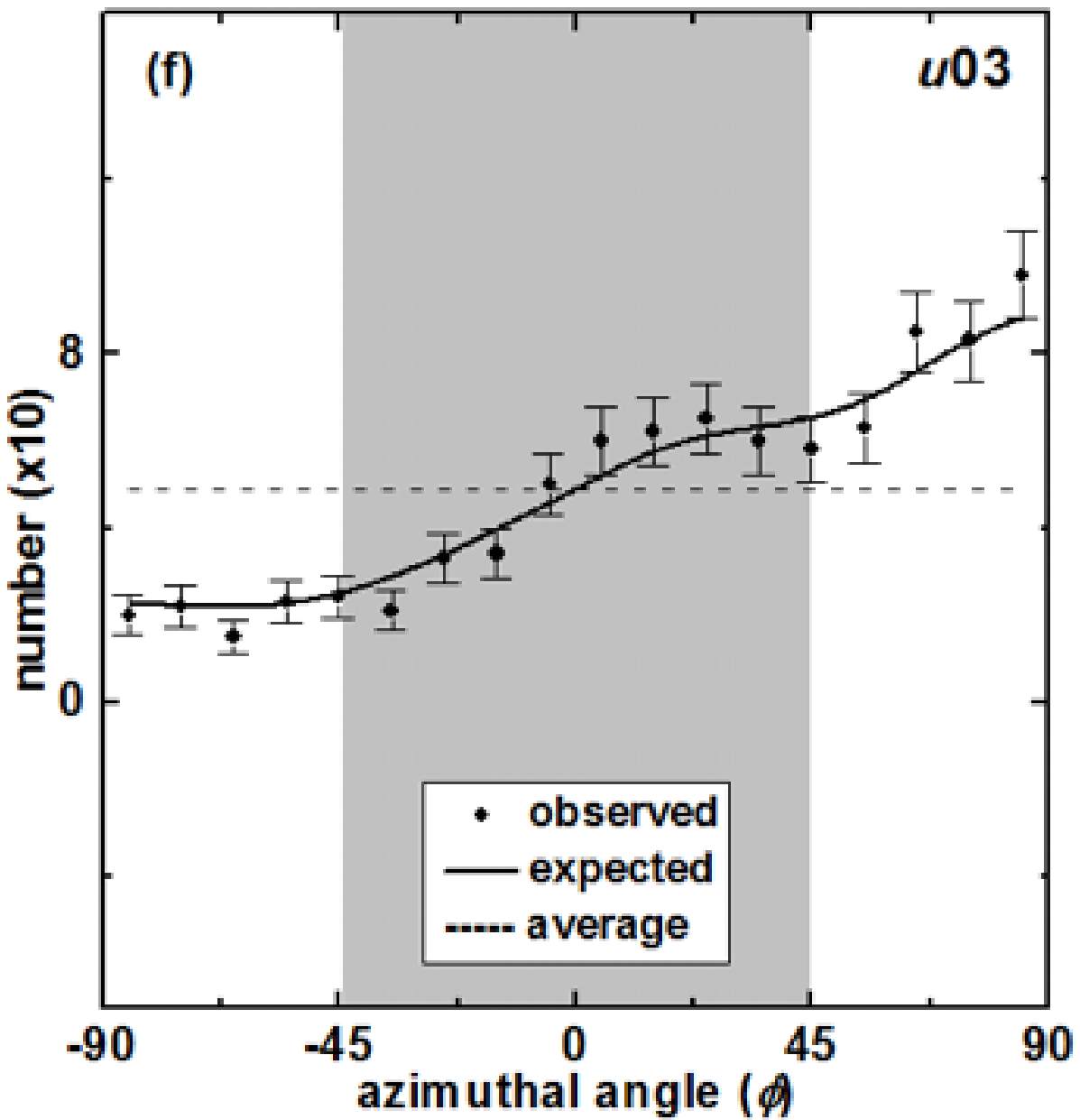}
\caption[]{The azimuthal ($\phi$) angle distributions of blue
shifted SDSS galaxies in 6 subsamples. The solid circles with
$\pm$ 1$\sigma$ error bars represent the observed distribution.
The solid line represents the expected isotropic distributions.
The average distributions (dashed) are shown for the comparison.}
\end{figure}
Fig. 4c shows the distribution of the spin vector projections of
high blue-shifted galaxies (subsample $i$03). Similar to the
subsample $i$02, all statistical parameters except P$(>\chi^2)$
show anisotropy. There is a significant hump at $+$90$^{\circ}$
and dips at $-$10$^{\circ}$ and 0$^{\circ}$. The SV projections of
high blue-shifted galaxies is found to be oriented tangentially
with respect to the center of the Milky Way.

All statistics except C/C($\sigma$) show anisotropy in the
$\phi$-distribution for subsample $u$01. In the histogram, several
humps and dips can be seen: humps at $-$90$^{\circ}$ and
$+$80$^{\circ}$, and dips in the region from $-$50$^{\circ}$ to
$+$50$^{\circ}$ (Fig. 4d). This subsample showed isotropy in the
polar and anisotropy in the azimuthal angle distribution. This
inconsistency is because of the inappropriate reference coordinate
system. A physical reference system should be identified in the
future.

Fourier test suggests isotropy in the subsample $u$02. There are
dips at $-$40$^{\circ}$, $+$20$^{\circ}$ and a hump at
$+$90$^{\circ}$ (Fig.4e). We observe weak anisotropy in the spin
vector projection of $u$02 galaxies with respect to galactic
coordinate system.

Interestingly, all statistics show isotropy in the distribution of
SV projections of high blue-shifted galaxies ($u$03). There are
small dips at 70$^{\circ}$ and 40$^{\circ}$, not significant to
alter the statistics of the subsample. There is a very good
agreement between the observed and expected distributions (Fig.
4f) Thus, a random orientation of spin vector projections is
found.

\subsection{Discussion}

In the numerous past literatures authors have studied the spatial
orientation of red-shifted galaxies in the field, clusters and
Superclusters and found mixed result: (1) noticed anisotropy
supporting either `pancake model' (Flin \& Godlowski 1986;
Godlowski 1993, 1994; Godlowski, Baier \& MacGillivray 1998, Flin
2001) as suggested by Doroshkevich (1973) or `primordial vorticity
model' (Baier, Godlowski \& MacGillivray 2003) as proposed by
Ozernoy (1978)(2) found isotropy suggesting a random orientation
of spin vectors of galaxies supporting `hierarchy model' (Bukhari
\& Cram 2003; Aryal \& Saurer 2005a, Aryal et al. 2006, 2012,
2013) as recommended by Peebles (1969).

In addition to these scenarios a bimodal tendency (Kashikawa \&
Okamura 1992), local anisotropy (Flin 1995, Djorgovski 1983, Aryal
\& Saurer 2004, 2005b), global anisotropy (Parnovsky, Karachentsev
\& Karachentseva 1994) are noticed. Godlowski \& Ostrowski (1999)
noticed a strong systematic effect, generated by the process of
deprojection of a galactic axis from its optical image. In
isolated Abell clusters, only brightest galaxies are
preferentially aligned (Trevese, Cirimele \& Flin 1992). Panko et
al. (2013), Godlowski (2012) and Godlowski et al. (2010) noticed a
dependence of alignment with respect to cluster richness.

The anisotropy is found mostly for those samples which is taken
from a limited region of the sky (e.g., clusters, sub-clusters,
groups, etc). For the field galaxies and superclusters, i.e., the
database taken from the large scale structure, authors notice a
random orientation supporting hierarchy model. In both the cases
(isotropy or anisotropy), a local effect is noticed. This effect
might arises due to the tidal connections between the neighboring
galaxies or because of the gravitational shearing effect.
Interestingly, blue-shifted galaxies support hierarchy model.
Thus, it can be concluded that the blue-shift is not different
cosmological effect, it is because of the peculiar velocity as
suggested by Aaronson et al. (1982).

\section{Conclusion}

We studied the spatial orientation of spin vectors of 5$\,$987
blue-shifted galaxies ($-$87.6 to $-$0.3 km$\,$s$^{-1}$) observed
through $i$- and $u$-filter.  These database were taken by SDSS
(7$^{th}$ data release). We classified our data into 6 subsamples
based upon their blue-shift. We have used the `PA-inclination'
method proposed by Flin \& Godlowski (1986) to convert two
dimensional observed parameters to three dimensional galaxy
rotation axes (polar and azimuthal angles) and carried out random
simulation by creating 5$\times$10$^6$ virtual galaxies in order
to remove selection effects from the database (Aryal \& Saurer
2000). To check for anisotropy or isotropy, we have carried out
three statistical tests: chi-square, auto-correlation and the
Fourier.

Since our observed distributions do not vary significantly from
the expected distribution, we have regarded the Fourier test as
more reliable. The local effects are examined by the correlation
test. We have taken $\chi^2$ test in order to check the binning
effect. In general, we found isotropic distribution of the spin
vector of galaxies in all six subsamples with respect to the
galactic coordinate system. There are humps and dips in the polar
angle distributions, and these humps and dips alter the
correlation test showing anisotropy in subsamples $i$02, $i$03,
and $u$03. Hence, local effect was observed in these subsamples
suggesting a local tidal connection between the rotation axes of
neighboring galaxies or the gravitational shearing effect.
However, in the rest of the subsamples ($i$01, $u$01, and $u$02)
small humps and dips do not alter the statistics of the total
subsample, so we suppose these as binning effects. In general, we
observe that there is no preferred alignment of spin vectors of
galaxies. Our results of $\theta$-distribution support the
hierarchical clustering scenario (Peebles 1969), which predicts
the random orientation of the directions of the spin vectors of
galaxies.

On the other hand, we found interesting results for the
$\phi$-distribution. As mentioned above, we have regarded the
Fourier test as more reliable. Out of six subsamples only two
subsamples ($i$01 and $u$03) show isotropy in all tests. Rest of
the subsamples ($i$02, $i$03,  $u$01, and $u$02) show anisotropic
distribution of the spin vector projections of galaxies. Also, all
subsamples (except $u$03) have negative value of the first order
Fourier coefficient $\Delta_{11}$. Therefore, in general the spin
vector projection of the galaxies tend to be orientated
perpendicular to the equatorial plane. However, this result needs
to be examined by using other reference system such as
Supergalactic coordinate system in the future.

\begin{acknowledgements}

This research has made use of the Sloan Digital Sky Survey (SDSS)
database (www.sdss.org) which is funded by Alfred P. Sloan
Foundation and the Participating Institutions. We acknowledge Dr.
Rajendra Parajuli, Amrit Campus, Kathmandu for his fruitful
comments on the manuscript. We thank Mr. R. K. Bachchan for
helping in the data compilation and Mr. S. Dangi for the data
reduction. One of the authors (SNY) acknowledges Central
Department of Physics, Tribhuvan University, Nepal for all kinds
of support for his Ph.D. work.

\end{acknowledgements}
%__________________________________________________________________________

\end{document}